\documentclass[amsmath,amssymb,amsbsy,prl,nofootinbib,twocolumn,tightenlines,superscriptaddress,floatfix]{revtex4-2}
\usepackage{amsfonts} 
\usepackage{amsmath} 
\usepackage{amssymb}
\usepackage{blindtext}
\usepackage{lineno}
\usepackage{cancel}
\usepackage{bm}
\usepackage{graphicx,nicefrac}
\usepackage{xfrac}
\usepackage[usenames, dvipsnames]{color} 
\usepackage{dcolumn}
\usepackage{epic} 
\usepackage{epsfig}
\usepackage{eucal} 
\usepackage{esdiff}
\usepackage{graphicx}
\usepackage[normalem]{ulem}
\usepackage{soul}
\usepackage{amsfonts} 
\usepackage{amsmath} 
\usepackage{amssymb}
\usepackage{physics}
\usepackage{graphicx,nicefrac}
\usepackage{blindtext}
\usepackage{lineno}
\usepackage{cancel}
\usepackage{scalerel}
\usepackage[normalem]{ulem}
\usepackage{xfrac}
\usepackage{bm}
\usepackage[usenames, dvipsnames]{color} 
\usepackage{dcolumn}
\usepackage{epic} 
\usepackage{epsfig}
\usepackage{eucal} 
\usepackage{esdiff}
\usepackage{graphicx}
\usepackage[breaklinks,colorlinks = true,linkcolor = magenta,urlcolor=magenta,citecolor=red]{hyperref}
\usepackage{soul}
\usepackage{subfigure}
\usepackage{textcomp}
\pagecolor{white}
\usepackage{verbatim}
\usepackage{wrapfig} 
\usepackage{xy} 
\usepackage{lineno}
\usepackage{float}
\usepackage{import}

\newcommand{\icts}{International Centre for Theoretical Sciences, Tata Institute of Fundamental Research, Bangalore 560089,India}
\newcommand{\JHU}{Department of Mechanical Engineering, Johns Hopkins University, Baltimore, Maryland 21218, USA}
\newcommand{\OIST}{Complex Fluids and Flows Unit, Okinawa Institute of Science and Technology Graduate University, 1919-1 Tancha, 
	Onna-son, Okinawa 904-0495, Japan}
\begin{document}
\title{Reduction of Triadic Interactions Suppresses Intermittency and Anomalous Dissipation in Turbulence}
\begin{abstract}
We investigate how the defining statistical features of three-dimensional turbulence respond to
systematic reductions of the Fourier-space triadic interaction network. Using direct numerical
simulations of both fractally and homogeneously decimated Navier--Stokes dynamics, we show that
progressive thinning of the set of active modes leads to a systematic suppression of intermittency
and, most strikingly, to the vanishing of the mean dissipation rate in the large-Reynolds-number
limit. Structure-function exponents collapse onto their dimensional values, the multifractal
singularity spectrum contracts, and the analyticity width extracted from the exponential spectral
tail increases monotonically with decimation---each indicating a substantial regularization of the
velocity field. Together, these results provide direct evidence that anomalous dissipation in
incompressible turbulence is not a generic property of the Navier--Stokes equations, but instead
requires the full combinatorial richness of their triadic nonlinear interactions.
\end{abstract}
	
	\author{Anikat Kankaria$^\clubsuit$}
	\email{anikat.kankaria@gmail.com}
	\affiliation{\icts}
	\author{Ritwik Mukherjee$^\clubsuit$}
	\email{ritwik.mukherjee@icts.res.in}
	\affiliation{\icts}
	\author{Sugan Durai Murugan} 
	\email{vsdmfriend@gmail.com}
	\affiliation{\JHU}
	\author{Marco Edoardo Rosti}
	\email{marco.rosti@oist.jp}
	\affiliation{\OIST}
	\author{Samriddhi Sankar Ray}
	\email{samriddhisankarray@gmail.com}
	\affiliation{\icts}
	\date{}
	\maketitle
	\def\thefootnote{$\clubsuit$}\footnotetext{These authors contributed equally to this work}\def\thefootnote{\arabic{footnote}}
	
	A central challenge in three-dimensional (3D), fully-developed
	turbulence research is to identify what aspects of the Navier--Stokes
	nonlinearity are indispensable for the emergence of multiscaling,
	intermittency, and anomalous
	dissipation~\cite{Frisch-Book,IshiharaARFM,Pandit2009,EyinkJFM2024}.  A
	possible candidate may well lie in the structure of the Fourier space
	triadic interactions between wavenumbers
	$(\mathbf{k},\mathbf{p},\mathbf{q})$ which satisfy
	$\mathbf{k}+\mathbf{p}+\mathbf{q}=0$ and whose collective, nonlinear
	action determines much of what makes the problem of turbulence
	difficult~\cite{Frisch-Book}.  It is tempting to ask if these key
	features of turbulence are robust properties of the three-dimensional
	(incompressible) Navier-Stokes equation, or whether the full
	combinatorial richness of the triadic interaction network is essential.
	More precisely, to what extent do intermittency, multifractality, and
	anomalous dissipation survive when the triadic network is progressively
	reduced? Related questions concerning the robustness of the dissipative
	anomaly itself have also been raised recently in direct numerical
	simulations of incompressible Navier--Stokes turbulence in periodic
	domains, where departures from the classical zeroth law of turbulence
	have been reported even in the absence of boundaries or explicit
	modifications of the equations~\cite{Iyer2025}.

	A direct way to address this question is to surgically reduce the set of active
	interactions by projecting the dynamics onto a prescribed subset of Fourier
	modes, as introduced by Frisch \textit{et al.}~\cite{Frischetal2012}.  
	Fourier decimation provides a particularly clean, isotropic, and statistically
	homogeneous procedure: By retaining each Fourier mode with a given probability, 
	one constructs a quenched projection that preserves
	incompressibility, the quadratic invariants, and the algebraic form of the
	nonlinear term, while systematically reducing the number of triadic interactions in either a fractal or homogeneous 
	fashion as we describe later.  Probes of this type --- including
	helicity-based surgeries~\cite{Biferale2012,BiferaleTiti2013} and geometric
	triad restrictions~\cite{Grossmannetal1996,Gurcan2012} --- have already demonstrated that the cascade and
	intermittency can be strongly modified by selective changes to the nonlinear
	couplings~\cite{Lanotte2015,Ray2015,Buzzicottietal2016Burgers,Buzzicotti2016NS,Ray2018,Picardoetal2020}. 
	
	Such a decimation procedure thins the triadic
	interaction network in controlled but complementary ways depending 
	on the exact nature of mode removal: The fractal version
	reduces the number of active modes scale by scale according to an effective
	dimension $D$ of the Fourier lattice~\cite{Frischetal2012}, while the 
	homogeneous version~\cite{Lanotte2015} removes modes uniformly across Fourier space. Neither procedure introduces any preferred physical direction or
	flow geometry, and both preserve incompressibility and the form of the
	nonlinear term. Taken together, they provide a systematic means of dismantling
	the interaction network, while keeping the underlying equations well defined, and hence 
	address the question posed earlier.
	
	In this paper we report results from direct numerical simulations
	(DNSs) of homogeneously and fractally decimated three-dimensional
	Navier-Stokes  dynamics. We find that, as the density of active Fourier
	modes is reduced, the flow becomes progressively consistent with the
	non-intermittent, simple scaling theory of
	Kolmogorov~\cite{Frisch-Book} accompanied by an \textit{almost}
	vanishing dissipation --- in contrast to anomalous dissipation in
	fully-developed turbulence~\cite{IshiharaARFM,EyinkJFM2024} --- with increasing Reynolds number.

	We begin with the incompressible, unit density 3D Navier-Stokes equation for the velocity field ${\bf u}$, with 
	its Fourier representation $\hat{\bf u}(\bf k,t)$, from which we define the decimated field 
	\begin{equation}
		\label{eq:decimOper}
		{\bf v}({\bf x},t)= {\cal P} \, {\bf u}({\bf x},t)=\hspace{-1mm}
		\sum_{{\bf k}}\hspace{-1mm} e^{i {\bf k \cdot \bf x}}\,\gamma_{\bf k}\hat{\bf u}(\bf k,t)\,,
	\end{equation}
	through the generalised Galerkin projector ${\cal P}$. 
	The quenched-in-time factors $\gamma_{\bf k}$ follow 
	\begin{equation}
		\label{eq:theta}
		\gamma_{\bf k} = 
		\begin{cases}
			1, & \text{with probability} ~h_k  \\
			0, & \text{with probability} ~1-h_k
		\end{cases}
		\;\; \text{with} ~k\equiv|{\bf k}|\,,
	\end{equation}
	and preserve Hermitian symmetry to ensure that ${\cal P}$ is a self-adjoint operator which 
	commutes with the nonlinear and viscous terms. 
	Such a micro surgery ensures that the invariants of the 
	original 3D problem --- namely energy and helicity --- are preserved~\cite{Frischetal2012,Ray2015}.
	
	\begin{figure*}
		\includegraphics[width = 1.00\linewidth]{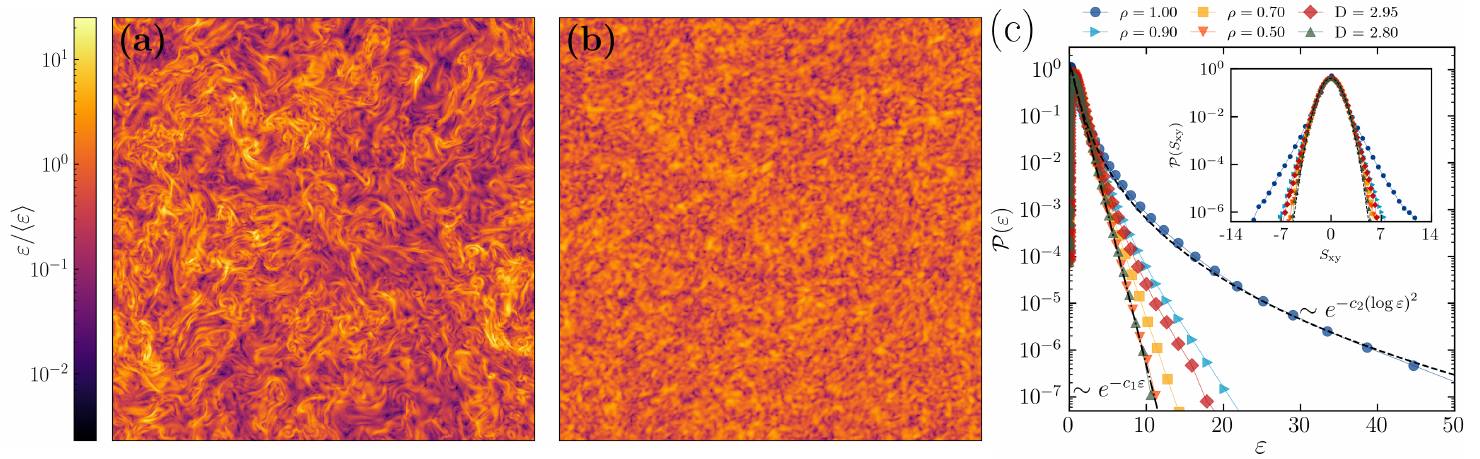}

		\caption{Pseudo-color plots of the energy dissipation field
		$\varepsilon$ on a two-dimensional slice for (a) the
		undecimated case $\rho=1$ and (b) a homogeneously decimated
		flow with $\rho=0.5$, shown on a logarithmic color scale.
		Decimation suppresses the intense filamentary structures
		characteristic of intermittent three-dimensional turbulence,
		resulting in a smoother dissipation field. (c) Probability
		density functions of $\langle \varepsilon \rangle$ for increasing decimation,
		showing a transition from log-normal to exponential tails.
		(Inset) PDFs of the longitudinal strain-rate component $S_{xy}$
		approaching a Gaussian form with increasing decimation.}
		\label{fig:dissp_slice}
	\end{figure*}

	This allows a simple reformulation of the Navier-Stokes equation, with 
	kinematic viscosity $\nu$ and pressure $p$, for the Fourier decimated velocity field 
	\begin{equation}
		\label{eq:decimNS}
		\frac{\partial {\bf v}}{\partial t} = {\cal P}[- {\bf \nabla}p - ({\bf v} \cdot {\bf \nabla}  {\bf v} )]  + \nu \,\nabla^2 {\bf v} +  {\bf F},
	\end{equation}
	with initial conditions ${\bf v}_0 = {\cal P} {\bf u}_0$. The projection 
	of the nonlinear term ensures that the evolution of the velocity field is restricted 
	on the subset of Fourier modes defined via Eqs.~\eqref{eq:decimOper}-\eqref{eq:theta}. Similarly, 
	the external force ${\bf F}$ which drives the flow to a statistically steady state 
	has the same compact support on the decimated Fourier lattice.

	There is freedom in the way we choose the factor $h_{k}$. In its
	original formulation --- the so-called \emph{fractal Fourier}
	decimation approach of Frisch \textit{et al.}~\cite{Frischetal2012} ---
	an effective dimensionality $D$ is obtained by choosing $h_k \propto
	(k/k_0)^{D-3}\,$, with $0< D \le 3\,$; the arbitrary wavenumber $k_0$
	can be set to unity.  This restriction of the dynamics on a
	$D$-dimensional Fourier sub-space results in an effective velocity
	field ${\bf v}$ with embedding dimension 3, but with degrees of freedom
	$\sim k^D$.  A second option ---  \emph{homogeneous} decimation --- is
	to decimate with $h_k = \rho,\quad \forall k$ with $0 < \rho \le
	1$~\cite{Lanotte2015,Ray2015,Buzzicottietal2016Burgers,Buzzicotti2016NS,Ray2018,Picardoetal2020}.
	Trivially, setting $D = 3$ or $\rho = 1$ leads to the
	usual three-dimensional Navier-Stokes equation.
	
	We solve the decimated Navier-Stokes equation on a 2$\pi$
	triply-periodic cubic box, with $512^3$ and $1024^3$ collocation points
	as well as different numerical solvers for consistency check.  Further,
	we drive the system to a statistically steady state using two different
	large-scale energy injections limited to the first two
	modes~\cite{Pope_forcing}, one that ensures a constant energy input and
	one that does not~\cite{eswaran_pope_1988a}. This allows us to achieve
	Taylor-scale Reynolds numbers $50 \le {\rm Re}_\lambda \le 550$, with
	${\rm Re}_\lambda = v_{\text rms}\lambda/\nu$; the mean kinetic energy
	dissipation rate $\langle \varepsilon \rangle = 2 \nu \langle S_{ij} S_{ij} \rangle$,
	where $S_{ij} = \tfrac{1}{2}(\partial_i v_j + \partial_j v_i)$ is the
	strain-rate tensor, and the root-mean-square velocity $ v_\text{rms}$
	allows us to define the Taylor micro-scale $\lambda^2 =  v^2_{\text rms} / \langle (\frac{\partial v}{\partial x} )^2 \rangle$, Finally, we use several different values of
	the fractal dimension $2.8 \le D \le 3.0$ and, for the homogeneous
	decimation, $0.4 \le \rho \le 1$; the simulations with $D = 3$ or $\rho
	= 1$ are solutions for the un-decimated Navier-Stokes equation, which
	serves as a reference state for many of the discussions that follow.
	
	\begin{figure} 
	\includegraphics[width = 1.00\linewidth]{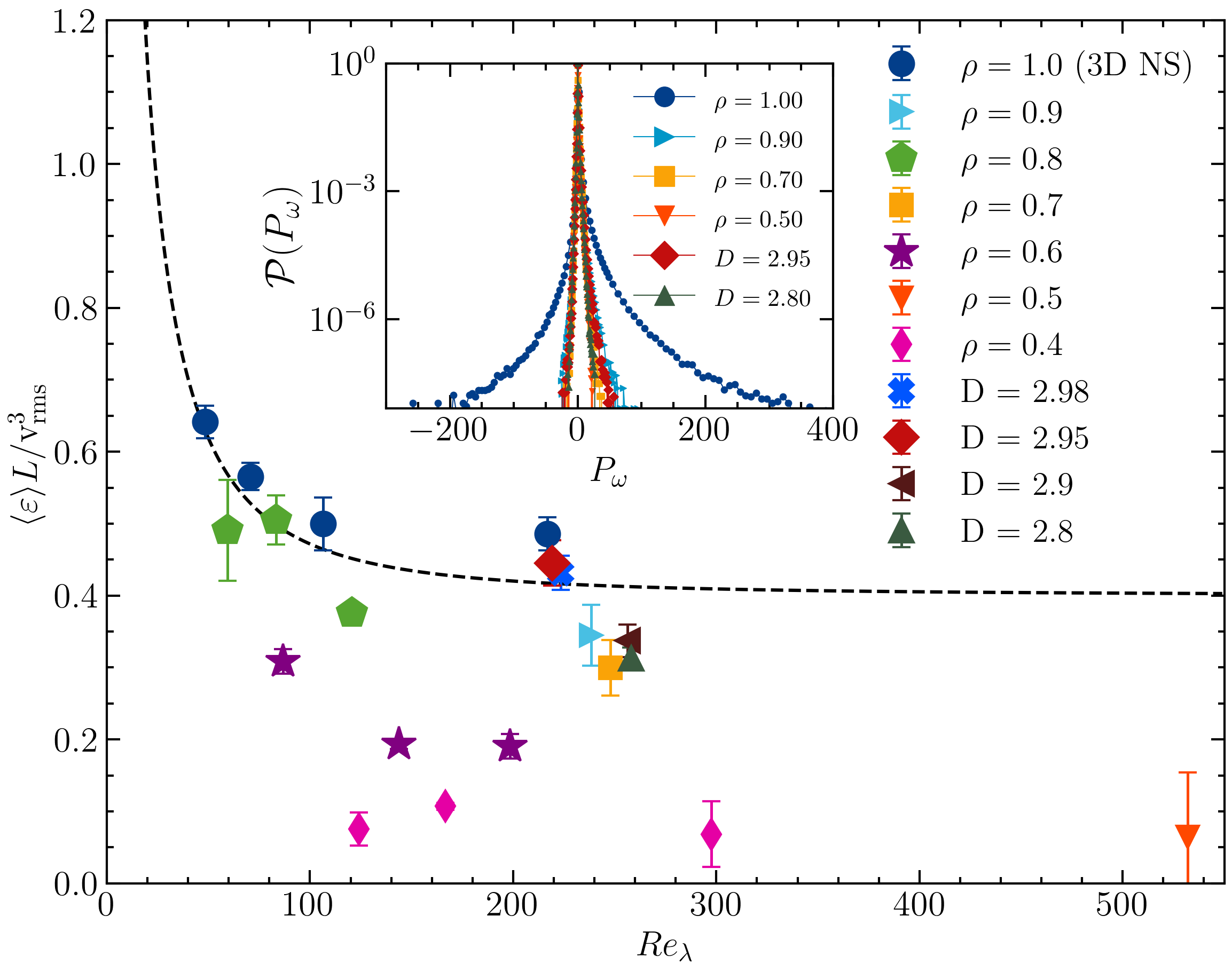} 
\caption{Mean dissipation rate $\varepsilon$ versus the Taylor-scale Reynolds
		number $Re_\lambda$ for different levels of Fourier decimation,
		compared with the undecimated three-dimensional Navier--Stokes
		case.  With increasing decimation, $\varepsilon$ decreases
		systematically with $Re_\lambda$, indicating a breakdown of
		dissipative anomaly. The black dashed line represent the parameterized dissipative anomaly curve~\cite{DONZIS_SREENIVASAN_YEUNG_2005,Cannon_Marco_2024}. (Inset) Probability density functions of
		the vorticity--strain interaction (vorticity production) for
		selected decimation levels, illustrating the suppression of
		extreme vortex-stretching events.  }
	\label{fig:anomaly} 
	\end{figure}
	
A useful insight on the role of triads comes from, for example, two-dimensional
slices of the energy dissipation rate $\varepsilon$~\cite{Picardoetal2020}.
This is illustrated, in Figs.~\ref{fig:dissp_slice}(a) and (b), by comparing
pseudocolor plots, on a logscale, of $\varepsilon$ from a fully
three-dimensional simulation ($\rho = 1$) to a highly decimated flow ($\rho =
0.5$), respectively. A comparison of these two figures show that decimation
leads to a suppression of the typical worm-like filaments,
characteristic of intermittent and fully-developed turbulence, and renders a
more featureless field devoid of structure~\cite{Picardoetal2020}. This absence
of intermittency suggests that the non-Gaussian tails of the strain field
$S_{ij}$ in turbulence should converge to a Gaussian with decimation. In the
inset of Fig.~\ref{fig:dissp_slice}(c) we show a representative plot of the
probability distribution function of $S_{\rm xy}$, with varying degrees of
decimation, including the $D = 3$ case, which confirms the intuition brought
out by Fig.~\ref{fig:dissp_slice}(b).  A further implication of such a trend is
the transition from a nominally log-normal distribution of the dissipation
field for $D = 3$~\cite{Frisch-Book} to one which is Chi-squared of 5th order
and hence exponentially distributed. We find compelling evidence of this in our
simulations as shown in Fig.~\ref{fig:dissp_slice}(c); this transition to an
exponential distribution was already noted by Picardo \textit{et
al.}~\cite{Picardoetal2020}. 

	Such a systematic suppression of intermittency naturally raises the question of 
	how the dissipative anomaly behaves under decimation, since the persistence of 
	finite mean dissipation as $\nu \to 0$ is, assumed to be, intimately tied to the 
	presence of strong gradients and intermittent fluctuations in the undiluted 
	three-dimensional flow. This motivates a direct examination of the mean 
	dissipation as a function of Reynolds number under controlled decimation, as we 
	now demonstrate.
	
	In Fig.~\ref{fig:anomaly} we show a plot of the normalised mean kinetic energy dissipation 
	$ \langle \varepsilon \rangle L/v^3_{\text rms}$, where the integral scale $L = \frac{\pi}{2 v^2_\textrm{rms}} \int_0^\infty \frac{E(k)}{k} dk$, 
	versus the Taylor-scale Reynolds number ${\rm Re}_\lambda$ for different 
	levels of decimation. The data for three-dimensional turbulence
	saturates to a finite, non-zero value as 
	${\rm Re}_\lambda \to \infty$ indicating dissipative anomaly and consistent 
	with a fitting function, as discussed in Refs.~\cite{DONZIS_SREENIVASAN_YEUNG_2005,Cannon_Marco_2024},
	which is indicated 
	by a black dashed line. 
	With increasing decimation, however, a clear trend emerges. Once the triadic network is 
	sufficiently tampered for $\rho \lesssim 0.5$ the mean dissipation decreases systematically with 
	${\rm Re}_\lambda$ and shows no sign of approaching a non-zero asymptote as marked by the black-dashed line.
	Instead, the data indicate that it is likely that $\langle \varepsilon \rangle L/v^3_{\text rms} \to 0$ as ${\rm Re}_\lambda \to \infty$ 
	for such a level of decimation is in clear violation of dissipative anomaly.
	
Further insight into the mechanism underlying Fig.~\ref{fig:anomaly} is provided by the inset,
which shows the probability density function of the vorticity--strain
interaction (often referred to as vorticity production); precise definitions of
the quantities involved, together with their geometric interpretation as well
as the associated $QR$ plots, are given in Appendix A. This interaction
governs vortex stretching in three-dimensional incompressible flows and is a
key ingredient of the forward transfer of kinetic energy to small scales. In
the undecimated Navier--Stokes dynamics, the distribution displays pronounced
non-Gaussian tails, reflecting the presence of rare but intense stretching
events that generate strong velocity gradients and sustain finite dissipation
at high Reynolds numbers. As the Fourier-space interaction network is
progressively decimated, these tails are systematically suppressed and the
distribution becomes increasingly concentrated. This trend indicates a
weakening of vortex stretching and, concomitantly, a reduced efficiency of the
nonlinear cascade in producing extreme small-scale activity. The inset thus
provides a dynamical counterpart to the behaviour seen in the main panel: the
decay of the mean dissipation rate with Reynolds number is accompanied by the
depletion of the very events responsible for feeding the cascade, driving the
flow toward a more regular regime in which anomalous dissipation is no longer
supported.

	To the best of our knowledge, this is the first direct demonstration of
	a violation of the dissipative anomaly in an incompressible
	Navier--Stokes system.  While our results do not address the origin of
	the anomaly in the fully three-dimensional case, they provide
	compelling evidence that its existence requires the full triadic
	structure of the Navier--Stokes nonlinearity. In particular, the
	suppression of $\varepsilon$ under decimation shows that removing even
	a statistically homogeneous subset of triads fundamentally alters the
	mechanism responsible for sustaining finite dissipation in the
	vanishing-viscosity limit.  Our results clearly demonstrate that the
	fine-scale structure of turbulence is highly sensitive to the
	connectivity of the triadic interaction network, even though the
	large-scale cascade and a power-law energy spectrum remain
	comparatively robust.  Viewed more broadly, these findings resonate
	with recent numerical evidence questioning the universality of the
	zeroth law of turbulence in incompressible Navier--Stokes flows
	\cite{Iyer2025}. In this context, our work identifies the combinatorial
	completeness of the triadic interaction network as a key structural
	ingredient required for sustaining anomalous dissipation in the
	infinite--Reynolds--number limit.
	
These observations connect naturally to rigorous constraints relating anomalous
dissipation to the behaviour of velocity increments.  Onsager’s criterion~\cite{EyinkSreenivasan2006} and
its mathematical refinements by Eyink~\cite{Eyink1994} and by Constantin, E,
and Titi~\cite{Constantin1994} place a critical H\"older threshold $h = 1/3$
for the possible existence of anomalous dissipation in the inviscid limit,
while the Duchon--Robert local energy balance~\cite{DuchonRobert2000} expresses
anomalous dissipation as a distributional limit of symmetric velocity
increments and ties it explicitly to third-order statistics.  Taken together,
these results imply that nonzero dissipation, in the infinite Reynolds number
limit, requires sufficiently singular velocity increments and nontrivial
third-order scaling, and they place restrictions on how rapidly the hierarchy
of structure functions may grow~\cite{Eyink2024}.

	Therefore, a further implication of this framework concerns the high-order exponents 
	themselves. Since anomalous dissipation demands a degree of singularity 
	incompatible with excessively smooth behaviour, the effective scaling ratios 
	$\zeta_p/p$ --- where $\zeta_p$ is the scaling exponent of the $p$-th order equal-time structure 
	function defined via $S_p(r) \equiv \langle 
	|\delta_r v|^p \rangle$, with $\delta_r v$ the longitudinal velocity increment between two points separated by an inertial 
	range distance $r$ --- for sufficiently large $p$ must lie strictly below the critical 
	value $1/3$ associated with energy conservation. Recent formulations of the 
	local energy balance make this explicit~\cite{EyinkJFM2024}: If $\langle \varepsilon \rangle$ is nonzero, then for suitable 
	$p>3$ one necessarily has $\zeta_p/p < 1/3$, provided the dissipation has even 
	mild spatial concentration. Thus, as Fourier decimation drives the flow toward 
	smoother behaviour and suppresses intermittency, one expects the ratios 
	$\zeta_p/p$ to drift upward toward the dimensional value $1/3$, signalling the 
	progressive loss—and eventual disappearance—of anomalous dissipation. This 
	connection provides a natural motivation for a direct examination of the 
	equal-time structure-function exponents under decimation, to which we now turn.
	
	\begin{figure}
		\includegraphics[width = 1.00\linewidth]{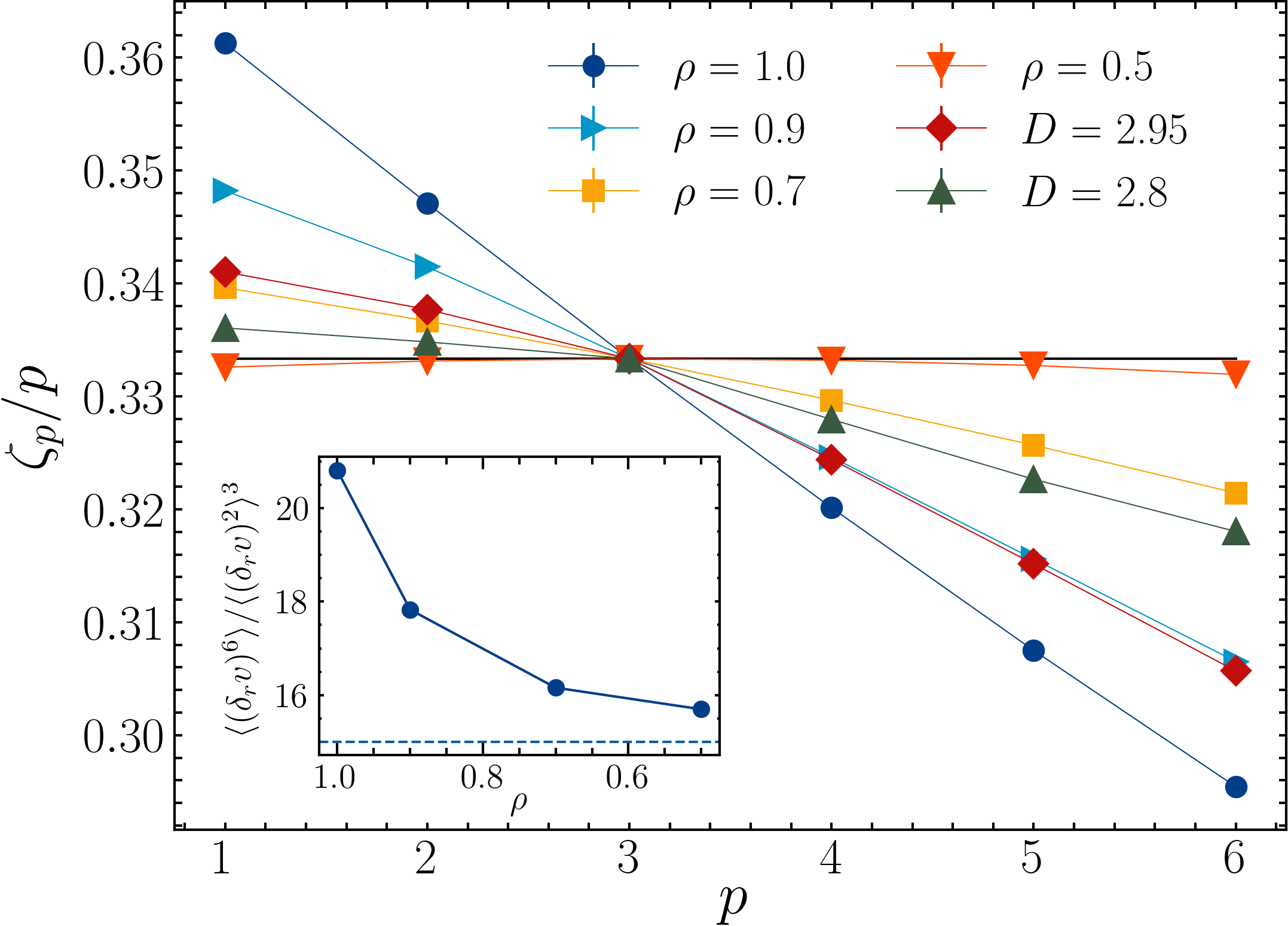} 
\caption{Ratios $\zeta_p/p$ of longitudinal equal-time structure-function
		exponents as a function of order $p$ for different levels of
		homogeneous and fractal Fourier decimation (see legend). The
		undecimated case exhibits clear departures from the dimensional
		prediction $\zeta_p/p = 1/3$, while increasing decimation leads
		to a progressive collapse toward the dimensional value. (Inset)
		Hyperflatness of the PDFs of longitudinal velocity increments at
		inertial-range separations, approaching the Gaussian value 15 with
		increasing decimation.}
		\label{fig:zetap}
	\end{figure}
	
	 Figure~\ref{fig:zetap} shows the 
	dependence of $\zeta_p/p$ on $p$, obtained via extended self-similarity~\cite{BenziESS,RayESS}, 
	for different values of $\rho$ and $D$, including, for reference the $\rho = 1$ (or $D = 3$) 
	three-dimensional problem. For the undecimated $D=3$ case, the exponents display the familiar 
	intermittency-induced departure from the dimensional line $1/3$. As the 
	decimation increases, however, the nonlinear curvature of the $\zeta_p/p$ 
	spectrum systematically diminishes: The high-order exponents move upward toward 
	their dimensional values, and the entire curve collapses progressively onto the 
	straight line $\zeta_p/p = 1/3$. This behaviour is fully consistent with the 
	regularity-based expectations discussed above and mirrors the concurrent 
	suppression of the dissipative anomaly. Furthermore a direct measurement of the hyperflatness 
	$\frac{\langle (\delta_r v)^6 \rangle}{\langle (\delta_r v)^2 \rangle^3}$ of the 
	probability density functions of the velocity increments show, as seen in the inset of Fig.~\ref{fig:zetap}, 
	a convergence to the Gaussian value 15 with increasing decimation.

\begin{figure}
		\includegraphics[width = 1.00\linewidth]{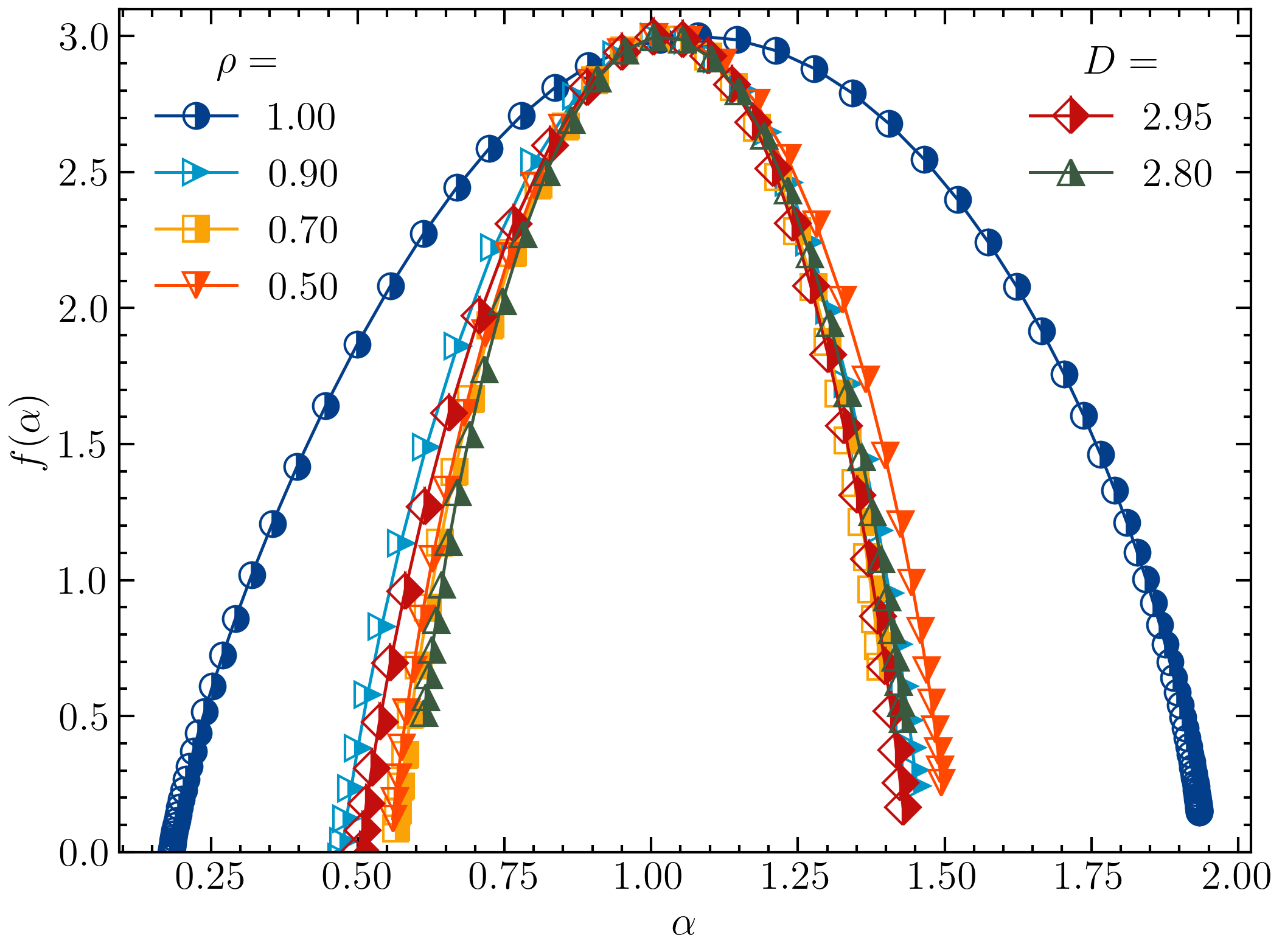}

\caption{Singularity spectra $f(\alpha)$ obtained from a multifractal analysis
		of the dissipation field for different levels of decimation.
		The undecimated case exhibits a broad spectrum, whereas
		increasing decimation leads to a systematic narrowing,
		indicating a progressive loss of multifractality.}

		\label{fig:singularity_spectrum}
	\end{figure}

The multifractal framework~\cite{,Frisch-Book,Frisch-Parisi} provides a natural bridge between the suppression of
intermittency under decimation and the behaviour of the dissipation field.
Since the dissipation is dominated by rare, intense gradient events, any
systematic smoothing of the flow---such as that produced by reducing the
connectivity of the triadic interaction network---must necessarily diminish the
range of local scaling exponents present in the velocity field. These are
commonly characterized in terms of the dissipation exponent $\alpha$, related
to the velocity H\"older exponent $h$ by $\alpha = 3h$, and their distribution
is encoded in the singularity spectrum
$f(\alpha)$~\cite{sreenivasan1991fractals,MS-Nucl,Mukherjeeetal2024}.  This
immediately implies that the singularity spectrum should contract as decimation
increases: Fewer extreme events correspond to a narrower spread of $\alpha$
values. This logical expectation is borne out clearly in the data. As shown in
Fig.~\ref{fig:singularity_spectrum}, the singularity spectra obtained from a
multifractal analysis of the dissipation field progressively narrow with
increasing decimation, demonstrating a collapse of multifractality and
confirming the direct link between the loss of intermittency and the
attenuation of anomalous dissipation. 

	The progressive narrowing of the singularity spectrum under decimation
	indicates a clear loss of fine-scale variability in the velocity field,
	suggesting that the flow becomes increasingly regular as the triadic
	interaction network is thinned. While the multifractal framework
	captures this through the contraction of $f(\alpha)$, it is equally
	valuable to complement this local, increment-based picture with a
	global diagnostic of smoothness. A natural candidate is provided by the
	analyticity-strip method, which quantifies the distance~$\delta$ of the
	closest complex singularity to the real axis through the exponential
	decay of the Fourier spectrum~\cite{Frisch-Book,Sulem-Sulem-Frisch,Muruganetal2020}. Because $\delta$ offers an independent,
	Fourier-space measure of regularity, examining its systematic variation
	with decimation provides a crucial cross-check of the multifractal
	results and further clarifies how the small-scale analytic structure of
	the flow is altered as the nonlinear couplings are reduced.
	
\begin{figure}
		\includegraphics[width = 1.00\linewidth]{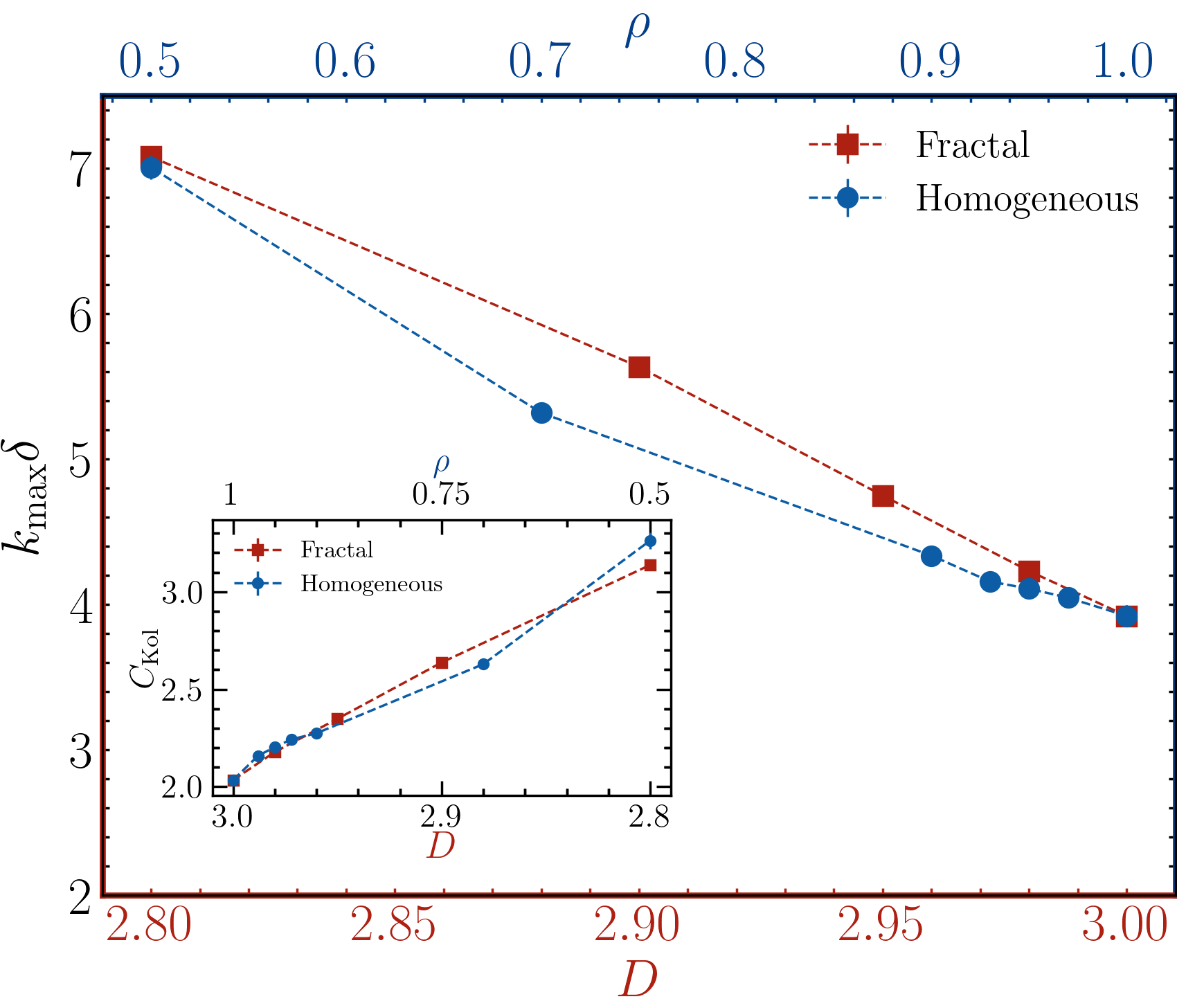}
\caption{Analyticity width $\delta$ as a function of the decimation parameter
		(either $D$ or $\rho$, see legend), extracted from the
		exponential decay of the statistically steady energy spectra.
		The monotonic increase of $\delta$ with decimation indicates a
		widening of the analyticity strip and a progressively smoother
		flow. (Inset) Kolmogorov constant as a function of decimation.}
		\label{fig:delta}
	\end{figure}

	To this end, we estimate the analyticity width $\delta$ for each
	decimation level by analysing statistically steady, shell-averaged
	energy spectra. In the high-wavenumber range where the spectrum
	exhibits exponential decay, the analyticity-strip framework predicts
	$\ln E(k) \approx -2\delta\, k + m\ln k + C$ so that $\delta$ may be
	extracted directly from the slope of $\ln E(k)$ versus
	$k$~\cite{Sulem-Sulem-Frisch}. For every decimated state, we compute
	the isotropic spectrum $E(k)$ over an ensemble of steady-state
	snapshots, identify the exponential tail through the plateau of the
	local slope $-\partial_k \ln E(k)$, and perform a weighted
	least-squares fit over this interval to obtain $\delta$, with
	uncertainties estimated by bootstrap resampling. The resulting
	dependence of $\delta$ on the decimation parameter, shown in
	Fig.~\ref{fig:delta}, reveals a clear and monotonic increase of the
	analyticity width as the fraction of active Fourier modes is reduced.
	This trend parallels the concomitant narrowing of the singularity
	spectrum and therefore provides strong, independent evidence for the
	progressive smoothing of the flow and the suppression of intermittency
	under decimation.  The inset of Fig.~\ref{fig:delta} shows a
	concomitant increase of the Kolmogorov constant $C_{\rm Kol}$,
	obtained from fits of the energy spectrum $E(k) = C_{\rm Kol}\,
	\varepsilon^{2/3} k^{-5/3}$, with decimation indicating that the
	reduced flux is accommodated by a renormalization of the inertial-range
	spectrum rather than a breakdown of Kolmogorov scaling~\cite{Frischetal2012}.
	
	In summary, our results demonstrate that the central signatures of
	fully developed three-dimensional turbulence --- anomalous dissipation,
	multiscaling of structure functions, and multifractality --- are not
	robust to a systematic reduction of the Fourier-space triadic
	interaction network. Both fractal and homogeneous decimation lead to a
	progressive weakening of intermittency, manifested in the collapse of
	the structure-function exponents toward their dimensional values, the
	marked narrowing of the multifractal singularity spectrum, and, most
	notably, the eventual disappearance of the dissipative anomaly. These
	observations provide direct, quantitative evidence that the persistence
	of finite dissipation in the vanishing-viscosity limit, along with
	attributes of fully developed turbulence depends, sensitively on the
	full combinatorial richness of the Navier--Stokes nonlinearity.

	\begin{acknowledgments}
		M.E.R. was supported by the Okinawa Institute of Science and Technology Graduate University (OIST) with subsidy funding from the Cabinet Office, Government of Japan. M.E.R. also acknowledges funding from the Japan Society for the Promotion of Science (JSPS), grant 24K17210 and 24K00810, and the computer time provided by the Scientific Computing \& Data Analysis section of the Core Facilities at OIST, and by HPCI, under the Research Project grants hp250021 and hp250035.
		SSR acknowledges  the CEFIPRA Project No 6704-1 for support and the hospitality of the 
		Okinawa Institute of Science and Technology (OIST) where parts of this work were carried out. The simulations were
		performed on the ICTS clusters Mario, Tetris, and Contra. AK, RM and SSR
		acknowledge the support of the Department of Atomic Energy, Government of India, under project
		no.RTI4019 and RTI4013.
	\end{acknowledgments}
	
	\bibliographystyle{apsrev4-2} 
	\bibliography{references}

\appendix 

\section*{Appendix A: QR diagnosis}

\begin{figure*}
\includegraphics[width = 1.00\linewidth]{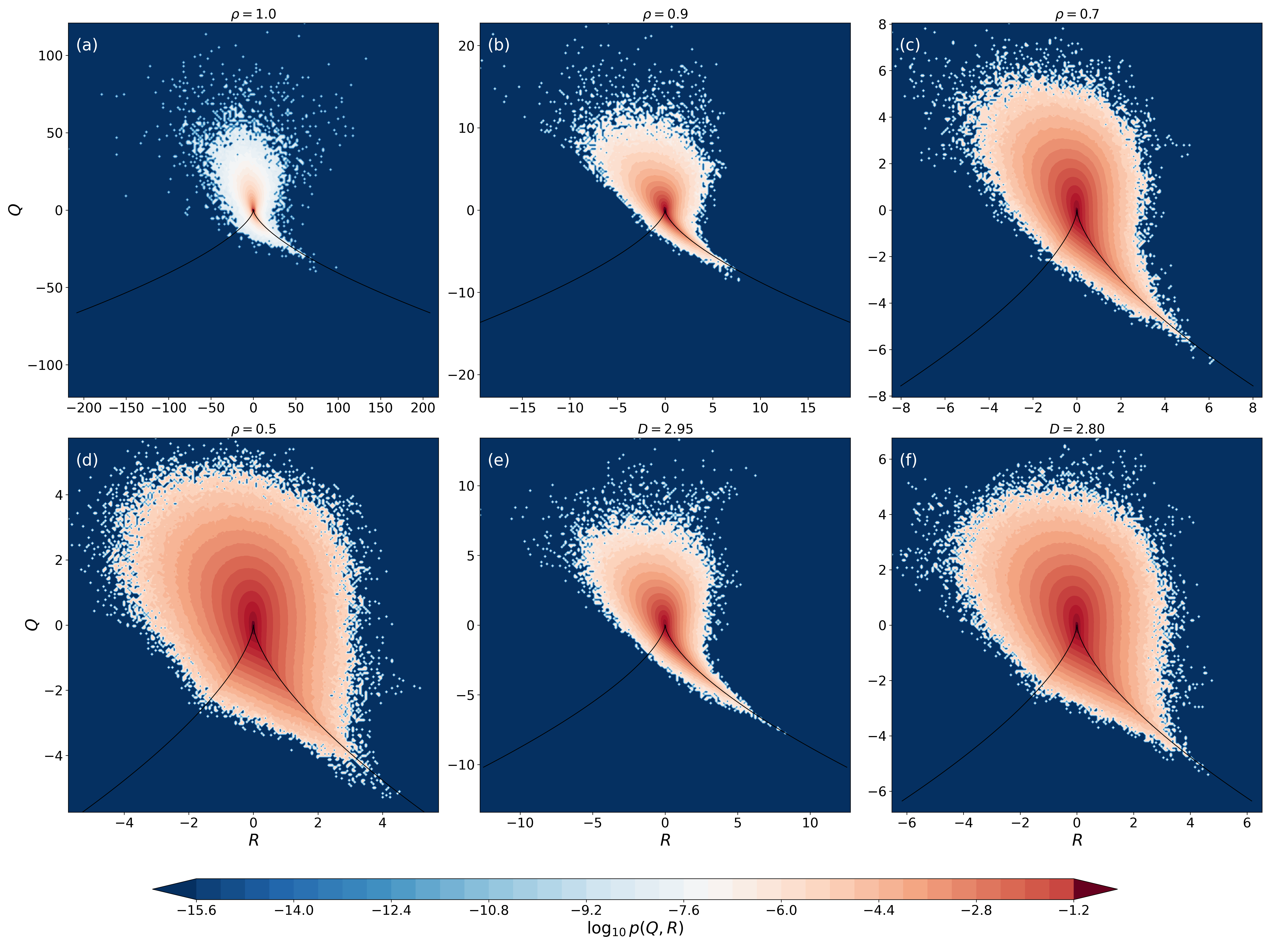}
\caption{Joint probability density functions of the second and third invariants
	$(Q,R)$ of the velocity-gradient tensor for different degrees of
	Fourier decimation (see labels). The undecimated Navier--Stokes case
	exhibits the characteristic teardrop-shaped distribution with broad
	support, reflecting intense small-scale velocity gradients and strong
	vortex stretching. As the level of decimation is increased, the support
	of the distribution contracts systematically and the accessible range
	of $Q$ and $R$ decreases, indicating a progressive suppression of
	extreme gradient events. This contraction of the $QR$ distributions
	provides clear geometric evidence that thinning the triadic interaction
	network kills small scales and drives the flow toward a smoother, more
	regular regime.}
\label{fig:QR}
\end{figure*}

A central quantity for characterizing small-scale dynamics in three-dimensional
incompressible turbulence is the vorticity--strain interaction, commonly
referred to as vorticity production. Let
\begin{equation}
\boldsymbol{\omega} = \nabla \times \boldsymbol{v}
\end{equation}
denote the vorticity field and
\begin{equation}
S_{ij} = \frac{1}{2}\left(\partial_i v_j + \partial_j v_i\right)
\end{equation}
the strain-rate tensor. The local rate of vorticity production is then given by
\begin{equation}
P_\omega = \omega_i S_{ij} \omega_j,
\end{equation}
which measures the stretching or compression of vorticity by the strain field.
Positive values of $P_\omega$ correspond to vortex stretching, a uniquely
three-dimensional mechanism responsible for the amplification of velocity
gradients and the forward cascade of kinetic energy to small scales. In fully
developed turbulence, the statistics of $P_\omega$ are strongly non-Gaussian,
reflecting the presence of rare but intense stretching events that play a
crucial role in sustaining intermittency and anomalous dissipation.

In the inset of Fig.~\ref{fig:anomaly} of the main text, we show the probability density
functions of $P_\omega$ for representative degrees of Fourier decimation. In
the undecimated Navier--Stokes dynamics, the distribution exhibits pronounced
non-Gaussian tails, reflecting the presence of rare but intense
vortex-stretching events that are responsible for generating extreme velocity
gradients. As the interaction network is progressively decimated, these tails
are systematically suppressed and the distribution becomes increasingly
concentrated around its mean. This clear depletion of extreme
vorticity-production events provides direct evidence that decimation weakens
the nonlinear mechanisms responsible for populating the smallest scales of the
flow.

To gain further geometric insight into the local flow topology underlying
vorticity production, it is useful to analyze the velocity-gradient tensor
$A_{ij} = \partial_j v_i$ through its second and third invariants, defined as
\begin{equation}
Q = -\frac{1}{2} \mathrm{Tr}(A^2), \qquad
R = -\frac{1}{3} \mathrm{Tr}(A^3).
\end{equation}

The joint probability distribution of $Q$ and $R$, commonly visualized in the
form of $QR$ plots, provides a compact representation of local flow
structures. In incompressible turbulence, the $(Q,R)$-plane is partitioned by
the discriminant curve
\begin{equation}
\Delta = \frac{27}{4} R^2 + Q^3 = 0,
\end{equation}
which separates regions dominated by rotational motion from those dominated by
strain. The characteristic teardrop-shaped distribution observed in the
undecimated Navier--Stokes dynamics is a robust statistical signature of
intense small-scale activity and strong vortex stretching.

Figure~\ref{fig:QR} shows the $QR$ plots for various degrees of homogeneous and
fractal Fourier decimation. In the undecimated case, the joint distribution
extends over a wide range of $Q$ and $R$, reflecting the presence of strong
velocity gradients and highly intermittent small-scale structures. As the level
of decimation is increased, however, the support of the distribution contracts
systematically. This contraction is accompanied by a marked reduction in the
range of accessible $Q$ and $R$ values, clearly visible through the progressive
narrowing of the axes in Fig.~\ref{fig:QR}.

This reduction in the extent of the $QR$ distributions provides direct evidence
that Fourier decimation suppresses the most intense small-scale events. By
thinning the triadic interaction network, decimation weakens vortex stretching,
limits the generation of extreme velocity gradients, and effectively removes
the tails of the velocity-gradient statistics. The shrinking of the $QR$ support
thus offers a geometric and statistical manifestation of the same
regularization observed in the dissipation field, vorticity-production
statistics, and structure-function scaling: decimation systematically kills
small scales, driving the flow toward a smoother, more regular regime in which
intermittency and anomalous dissipation can no longer be sustained.

\end{document}